\documentclass[nofootinbib]{revtex4}
\pdfoutput=1  %necessary for JHEP
\usepackage[latin1]{inputenc}
\usepackage{ae,aecompl}
\usepackage{amsmath,amssymb,mathrsfs}
\usepackage{graphicx}
\usepackage[]{units}
\usepackage[usenames,dvipsnames]{color} % p usar nomes de cores
\usepackage{hyperref}
\hypersetup{
   colorlinks=true,       % false: boxed links; true: colored links
    linkcolor=Blue,          % color of internal links
    citecolor=Plum,        % color of links to bibliography
    filecolor=magenta,      % color of file links
    urlcolor=YellowOrange           % color of external links
}

\providecommand{\cp}{\mathsf{CP}}

\providecommand{\om}{\omega}

\providecommand{\cpmutau}{\cp^{\mu\tau}}

\providecommand{\xlink}[1]{\href{http://arxiv.org/abs/#1}{#1}}
\providecommand{\eq}[1]{\begin{equation} #1 \end{equation}}
\providecommand{\eqali}[1]{\begin{equation}\begin{aligned} #1
    \end{aligned}\end{equation}}
\DeclareMathOperator{\im}{\mathrm{Im}}
\DeclareMathOperator{\re}{\mathrm{Re}}
\providecommand{\ZZ}{\mathbb{Z}}
\providecommand{\mtrx}[1]{\begin{pmatrix} #1 \end{pmatrix}}
\providecommand{\mss}[1]{\mbox{\scriptsize $#1$}}
\providecommand{\tp}{{\mss{\mathsf{T}}}}
\DeclareMathOperator{\diag}{\mathrm{diag}}
\providecommand{\ums}[2][1]{\ml{\tfrac{#1}{#2}}}
\providecommand{\ml}[1]{\mbox{\large $#1$}}
\providecommand{\mfn}[1]{\mbox{\footnotesize $#1$}}
\providecommand{\aver}[1]{\langle #1 \rangle}

\begin{document}
%%%%%%%%%%%%%%%%%%%%%%%%%%%%%%%%%%%%%%%%%%%%%%%%%
\title{
Mu-tau reflection symmetry with a texture-zero
}
\author{C.~C.~Nishi}
\email{celso.nishi@ufabc.edu.br}
\affiliation{
Centro de Matemática, Computação e Cognição\\
Universidade Federal do ABC - UFABC, 09.210-170,
Santo André, SP, Brazil
}

\author{ B. L. S\'anchez-Vega}%
\email{brucesan@ift.unesp.br}
\affiliation{
	Instituto  de F\'\i sica Te\'orica -- Universidade Estadual Paulista \\
	R. Dr. Bento Teobaldo Ferraz 271, Barra Funda\\ S\~ao Paulo - SP, 01140-070,
	Brazil
}

%\date{\today}
%%%%%%%%%%%%%%%%%%%%%%%%%%%%%%%%%%%%%%%%%%%%%%%%%
\begin{abstract}
The $\mu\tau$-reflection symmetry is a simple symmetry capable of predicting all the 
unknown CP phases of the lepton sector and the atmospheric angle but too simple to 
predict 
the absolute neutrino mass scale or the mass ordering.
We show that by combining it with a discrete abelian symmetry in a nontrivial way 
we can additionally enforce a texture-zero and obtain a highly predictive scenario 
where
the lightest neutrino mass is fixed to be in the few meV range for two normal ordering (NO)
solutions or in the tens of meV in one inverted ordering (IO) solution.
The rate for neutrinoless double beta decay is predicted to be negligible for NO or 
have effective mass $m_{\beta\beta}\approx 14\text{ -- }29\,\unit{meV}$ for IO, right in the region
to be probed in future experiments.
\end{abstract}
%%%%%%%%%%%%%%%%%%%%%%%%%%%%%%%%%%%%%%%%%%%%%%%%%
% \pacs{12.60.Fr, 11.30.Er, 11.30.Fs, 11.30.Qc}
% 14.80.Ec 	Other neutral Higgs bosons
% 14.80.Fd 	Other charged Higgs bosons 
% 11.30.Fs 	Global symmetries (e.g., baryon number, lepton number) 
% 11.30.Qc 	Spontaneous and radiative symmetry breaking 
% \keywords{Higgs, multi-Higgs models, 2HDM, global symmetry, custodial symmetry,
% symmetry breaking}
%\twocolumn
\maketitle
%\tableofcontents
%%%%%%%%%%%%%%%%%%%%%%%%%%%%%%%%%%%%%%%%%%%%%%%%%
\section{Introduction}
\label{sec:intro}

After the discovery of the nonzero value of the reactor angle $\theta_{13}$ in 2012\,\cite{dayabay}, 
a few unknowns remain in the neutrino sector if neutrinos are Majorana: 
the ordering of neutrino masses, the absolute scale of neutrino masses and 
the values of the three CP phases -- one of Dirac type and two of Majorana
type.
Major experimental efforts on neutrino oscillations are now focused on determining 
the Dirac CP phase and the mass ordering.

One of the simplest symmetries that can predict all the CP phases and yet allow CP 
violation is the symmetry known as $\mu\tau$-reflection symmetry or 
$\cpmutau$ symmetry where the neutrino sector is invariant by exchange of the muon
neutrino with the tau antineutrino\,\cite{mutau-r}.
This symmetry ensures that the Dirac CP phase is maximal ($\delta=\pm 90^\circ$) while the Majorana phases are 
trivial allowing the possibility of discrete choices of the CP parities.
Additionally, the atmospheric angle $\theta_{23}$ is predicted to be maximal 
($45^\circ$) at the same time that $\theta_{13}$ is permitted to be nonzero.
These values for the neutrino parameters are still allowed by current global fits
and in fact there are hints that $\delta\sim -90^\circ$\,\cite{schwetz,lisi}.
The fixed values for the CP phases also lead to characteristic bands for the possible 
effective mass of neutrinoless double beta decay but still allows leptogenesis 
to occur if flavor effects are taken into account\,\cite{cp.mutau}.

It was shown in Ref.\,\cite{cp.mutau} that the simplest way of implementing 
$\cpmutau$ and guarantee diagonal charged lepton masses is to combine $\cpmutau$ 
with the combination $\mathtt{L}_\mu-\mathtt{L}_\tau$ of family lepton numbers. 
This combination is trivial in the sense that these two symmetries commute and this 
feature allows us to avoid the vev alignment problem that requires special 
treatment in many models with discrete nonabelian flavor symmetries\,\cite{review}. 
In fact, $\cpmutau$ can be successfully implemented, sometimes accidentally, in 
models with discrete flavor symmetries\,\cite{mutau:r:models}. More recently, it 
was shown that maximal $\theta_{23}$ and maximal Dirac CP phase can be obtained 
without the explicit imposition of a CP symmetry\,\cite{real.sym} at the expense of 
requiring vev alignment and losing the predictions for the Majorana phases. See 
Ref.\,\cite{mutau:review} for a review on $\cpmutau$
and also on the $\mu\tau$ interchange symmetry\,\cite{mutau}.

Our main goal here is to show that one can have $\cpmutau$ symmetry along with a 
discrete abelian symmetry that ensures a \textit{one-zero texture}\,\footnote{%
Due to $\cpmutau$ symmetry some entries are related and only one-zero textures are 
allowed so that none of these cases correspond to the two-zero textures studied 
first in Ref.\,\cite{frampton}. }. This reduces the number of free parameters in 
the neutrino mass matrix from five to four to account for the four observables 
$\Delta m^2_{21},\Delta m^2_{32},\theta_{12},\theta_{13}$ -- the rest are fixed from symmetry --
and we obtain a highly predictive 
scenario where the absolute neutrino mass is fixed and further correlations of 
parameters appear. 

Our approach is a combination of two very different approaches to lepton flavor:
(a) texture-zeros that increase predictivity and relate mixing angles with masses\,\cite{frampton,2-zero} and (b) symmetries that fully or partly determine the mixing structure independently of the masses \cite{mass.independent,review}.
The former usually requires abelian symmetries\,\cite{tanimoto} while the latter requires
nonabelian discrete symmetries\,\cite{review}.
The most general cases of texture-zeros has been analyzed recently in Ref.\,\cite{grimus:texture}
where generic texture-zeros are required for the mass matrices of both charged leptons and neutrinos.
These cases include the well-studied parallel structures where both mass matrices have the same texture-zeros\,\cite{parallel}.
We refer to Ref.\,\cite{texture:rev} for a review.
In contrast, residual CP symmetries have also been considered to determine 
the mixing angles (and CP phases)\,\cite{holthausen,remnant.cp,cp.zero}.

If we give up symmetries that predict mixing angles, it is also possible to use 
nonabelian flavor symmetries to obtain texture-zeros together with equal elements 
in the neutrino mass matrix, a scenario known as hybrid texture\,\cite{hybrid}, 
recently generalized in Ref.\,\cite{cebola}. Nonabelian groups are generally 
required because one needs noncommuting symmetries.
The combination of two noncommuting 
symmetries cannot be arbitrary and needs to fulfill some compatibility rules so 
that 
the whole group closes.\footnote{%
It is not difficult to construct infinite discrete groups when combining noncommuting 
symmetries with three dimensional representations\,\cite{hernandez}.
}
In fact, some consistency conditions are required to combine a CP symmetry with a nonabelian 
discrete flavor group\,\cite{holthausen}.
For this reason, we choose the simplest setting where we combine an abelian discrete symmetry   
with $\cpmutau$ in a consistent but nontrivial way. 
As a result, the chosen abelian symmetry will simultaneously be responsible for the 
diagonal charged lepton masses and for the texture-zero in the neutrino mass matrix.

The outline of this work is as follows: in Sec.\,\ref{sec:sym} we show the two symmetries
that will be combined in a consistent way.
A useful parametrization of the neutrino mass matrix is shown in Sec.\,\ref{sec:param}
and the possible one-zero textures that are compatible with data are presented in Sec.\,\ref{sec:zero}.
Section \ref{sec:model} develops an example model and our conclusions can be read in
Sec.\,\ref{sec:conclusion}.

%%%%%%%%%%%%%%%%%%%%%%%%%%%%%%%%%%%%%%%%%%%%%%%%%
\section{Underlying symmetry}
\label{sec:sym}

We say the neutrino mass matrix is invariant by $\cpmutau$ symmetry, or $\mu-\tau$ 
reflection\,\cite{mutau-r}, when
\eq{
\label{cp.mutau}
M_\nu=
\left(
\begin{array}{ccc}
 a & d & d^* \\
 d & c & b \\
 d^* & b & c^* \\
\end{array}
\right)\,,\quad
\text{with real $a,b$ and $\im(d^2 c^*)\neq 0$}\,.
}
By rephasing we can eliminate either the phase of $c$ or $d$ so that we have five 
real continuous parameters in total.
This form for the neutrino mass matrix in the flavor basis (diagonal charged lepton 
masses) is known to predict maximal $\theta_{23}=45^\circ$ and $\delta=\pm 
90^\circ$ at 
the same time that $\theta_{13}\neq 0$ is allowed\,\cite{mutau-r}.
Additionally, the Majorana phases are trivial and four discrete choices for the CP 
parities are possible\,\cite{mutau-r,cp.mutau}.
These features lead to characteristic predictions for the neutrinoless double beta 
decay rate and leptogenesis\,\cite{cp.mutau}.

The mass matrix \eqref{cp.mutau} has five real independent parameters to describe 
five observables: $\theta_{12},\theta_{13},m_1,m_2,m_3$. One of them -- the 
absolute 
neutrino mass scale -- is unknown.
Given the same number of parameters and observables, there is no sharp 
prediction for the latter if only $\cpmutau$ is present.
We will show in the following that one can have $\cpmutau$ symmetry along with an 
abelian symmetry that ensures a \textit{one-zero texture}.
With one less parameter we obtain a definite prediction on the absolute neutrino 
mass.
It is clear that $d$ or $c$ cannot vanish because the resulting matrix after 
appropriate rephasing is symmetric by $\nu_\mu-\nu_\tau$ interchange which leads to 
the experimentally excluded value of $\theta_{13}=0$.

In order to implement $\cpmutau$ naturally, it was shown in Ref.\,\cite{cp.mutau} that the only way we can combine a residual 
$U(1)$ symmetry in the charged lepton sector and a residual CP symmetry in the 
neutrino sector with nontrivial CP violation is to consider
a $U(1)$ generated by the combination of lepton flavor numbers 
$\mathtt{L}_\mu-\mathtt{L}_\tau$ and $\cpmutau$ as the CP symmetry.
In group theoretical terms, other combinations such as $\mathtt{L}_e-\mathtt{L}_\mu$ 
and $\cp^{e\mu}$ are allowed but they are not phenomenologically viable.

If we allow the electron flavor to have nontrivial charge and consider $\ZZ_n$ instead of 
$U(1)$, other possibilities arise beginning with $\ZZ_8$\,\cite{cp.mutau}.
Here we use such a possibility to ensure $\cpmutau$ symmetry with a texture-zero.
We assign $\ZZ_8$ charges to the charged leptons $(e,\mu,\tau)$ as follows
\eq{
\label{def:T}
T=\mtrx{-1&&\cr &\om_8&\cr &&\om_8^3}\,,
~~\om_8=e^{i2\pi/8}.
}
This symmetry ensures diagonal charged lepton masses.
In contrast, the $\cpmutau$ symmetry acts as usual on the left-handed neutrino fields 
$\nu_{\alpha L},~\alpha=e,\mu,\tau$, as
\eq{
\label{cp:nu}
\cpmutau:\quad\nu_{\alpha L}\to X_{\alpha\beta}\nu_{\beta L}^{cp}\,,
}
where $cp$ denotes the usual CP conjugation and $X$ is $\nu_\mu$-$\nu_\tau$ interchange,
\eq{
X=\mtrx{1&0&0\cr0&0&1\cr0&1&0}\,.
}

We can think that these two symmetries -- $\ZZ_8$ generated by $T$ and $\cpmutau$ 
-- initially act on the left-handed lepton doublets $(L_e,L_\mu,L_\tau)$ before 
they are spontaneously broken.
Then the two symmetries act on the same space and $\cpmutau$ induces the 
following automorphism on $\ZZ_8$\,\cite{holthausen}:
\eq{
T\to XT^*X^{-1}=T^5\,.
}
We also note that the rephasing transformations that preserve $\ZZ_8$ in 
\eqref{def:T} and $\cpmutau$ in \eqref{cp:nu} are of the form
\eq{
\label{rephasing}
L_e\to \pm L_e,\quad
L_\mu\to e^{i\alpha}L_\mu,\quad
L_\tau\to e^{-i\alpha}L_\tau.
}
It is clear that these transformations also preserve the form of the mass matrix in 
\eqref{cp.mutau} and can be used to make $c$ or $d$ real.
Flavor independent rephasing by $i$ also preserves the form of the mass matrix 
(flips the sign of $a,b$) but changes $\cpmutau$ by a global sign.
Hence, only the relative sign of $a$ and $b$ is significant.

On the other hand, each quadratic combination $\bar{\nu}^c_{\alpha L}\nu_{\beta L}$ 
(Majorana neutrinos) that will give rise to the 
neutrino mass matrix carries the following $\ZZ_8$ charges:
\eq{
\label{Z8:mnu}
\bar{\nu}^c_{\alpha L}\nu_{\beta L}\sim 
% (M_\nu)_{\alpha\beta}\sim 
\mtrx{1 &\omega_8^5 &\omega_8^{-1}\cr
	\star & \omega_8^2 & -1\cr
	\star & \star & \omega_8^{-2}\cr
	}\,.
}
As all entries carry different charges (including the trivial), we can arrange the appropriate 
texture-zero in the $(ee)$ or $(\mu\tau)$ entry by making the nonzero entries come from 
the vacuum expectation values of scalars carrying the desired quantum 
numbers\,\cite{tanimoto}. We give an explicit construction in Sec.\,\ref{sec:model}.

%%%%%%%%%%%%%%%%%%%%%%%%%%%%%%%%%%%%%%%%%%%%%%%%%
\section{Parametrization}
\label{sec:param}

It was shown in Ref.\,\cite{mutau-r} that the $\cpmutau$ symmetric matrix 
\eqref{cp.mutau} can be diagonalized by a matrix of the form
\eq{
\label{U0}
U_0=
\mtrx{u_1 & u_2 & u_3\cr
    w_1 & w_2 & w_3\cr
    w_1^* & w_2^* & w_3^*
    }\,,
}
with $u_i$ real and conventionally positive.
The diagonalization performs
\eq{
\label{diag}
U_0^\tp M_\nu U_0=\diag(m_1',m_2',m_3')\,,
}
where $m_i'=\pm m_i$, with $m_i$ being the neutrino masses.
Therefore, the full diagonalizing matrix can be written as
\eq{
U_\nu=U_0K\,,
}
where $K$ is a diagonal matrix of $1$ or $i$ depending on the sign on 
\eqref{diag}.
We can classify the cases according to the sign of $m_i'$ or the diagonal entries 
of $K^2$\,\cite{cp.mutau} as
\eq{
(+++),~(-++),~(+-+),~(++-)\,.
}
There is also the freedom to replace $U_0$ by $U_0^*$ in \eqref{diag}, together 
with $M_\nu\to M_\nu^*$. 
This replacement flips the sign of the Dirac CP phase and the Jarlskog invariant, 
leaving the rest of observables invariant. 

Comparing \eqref{U0} to the standard parametrization of the PMNS matrix and choosing the convention 
that $-iw_3>0$ we arrive at the parametrization
\eq{
\label{U0:param}
U_0=\left(
\begin{array}{ccc}
 1 & 0 & 0 \\
 0 & \frac{1}{\sqrt{2}} & \frac{\pm i}{\sqrt{2}} \\
 0 & \frac{1}{\sqrt{2}} & \frac{\mp i}{\sqrt{2}} \\
\end{array}
\right)
\mtrx{ c_{13}&0&s_{13}\cr
    0&1&0\cr
    -s_{13}&0&c_{13}}
\mtrx{ c_{12}&s_{12}&0\cr
    -s_{12}&c_{12}&0\cr
    0&0&1}
\,. 
} 
The $\pm$ sign coincides with the Dirac CP phase given by $e^{i\delta}=\pm i$ and $\theta_{23}=45^\circ$ is 
fixed by symmetry. Note that the standard parametrization corresponds to 
$\diag(1,1,-1)U_0\diag(1,1,\mp i)$.

If we invert the relation \eqref{diag} by using \eqref{U0:param} with the top 
signs, we can write the parameters $a,b,c,d$ in terms of the neutrino masses and 
mixing angles: 
\eqali{ 
\label{params}
a&=c^2_{13}(m_1'c^2_{12} +m_2's^2_{12})+m_3' s^2_{13}\,, 
\cr 
b&=\ums{2}\big[m_1's^2_{12}+m_2'c^2_{12} +s^2_{13}(m_1'c^2_{12}+m_2's^2_{12}) 
+m_3' c^2_{13}\big]\,, 
\cr 
d&=\frac{c_{12}s_{12}c_{13}}{\sqrt{2}}(m_2'-m_1') 
+i\frac{s_{13}c_{13}}{\sqrt{2}}\big[-m_3'+m_1'c^2_{12}+m_2's^2_{12}\big]\,, 
\cr 
c&=\ums{2}\big[m_1'(s^2_{12}-c^2_{12}s^2_{13})+m_2'(c^2_{12}-s^2_{12}s^2_{13})
-m_3'c^2_{13}\big] 
+i\,c_{12}s_{12}s_{13}(m_2'-m_1')\,. 
} 
Choosing the bottom signs in \eqref{U0:param} corresponds to taking $d\to d^*$ and
$c\to c^*$.
The phases of $c,d$ are also convention dependent as they can be transferred from one to 
the other by the rephasing transformation \eqref{rephasing}.
A rephasing invariant CP-odd quantity is 
\eq{
\im(c^*d^2)=\pm\ums{2}(m_1'-m_2')(m_2'-m_3')(m_3'-m_1')s_{13}c^2_{13}s_{12}c_{12}\,.
}
for both signs in \eqref{U0:param}.
This invariant is clearly nonzero for physical values and corresponds to one of the 
invariants in Ref.\,\cite{smirnov:inv} adapted to the $\cpmutau$ symmetry case.

We can also note that if we perform a change of basis of $M_\nu$ only by the first 
matrix of \eqref{U0:param}, we obtain a real symmetric matrix which can be 
diagonalized by a real orthogonal matrix.
If we compare the trace of $M_\nu$ and $M_\nu^2$ in this new basis as well as the 
determinant, we obtain the following relations:
\eqali{
m_1'+m_2'+m_3'&=a+2b\,,
\cr
m_1^2+m_2^2+m_3^2&=a^2+2(b^2+|c|^2+2|d|^2)\,,
\cr
m_1'm_2'm_3'&=a(b^2-|c|^2)-2b|d|^2+2\re(c^*d^2)\,.
}
This means that we can trade three among $a,b,|c|,|d|,\im(c^*d^2)$ by the three 
neutrino masses $m_i$ for each choice of CP parities.

%%%%%%%%%%%%%%%%%%%%%%%%%%%%%%%%%%%%%%%%%%%%%%%%%
\section{Possible one-zero textures}
\label{sec:zero}

A texture-zero in the $(ee)$ or $(\mu\tau)$ entries of the $\cpmutau$ symmetric 
neutrino mass matrix \eqref{cp.mutau} is possible depending on the neutrino CP 
parities and the mass ordering. 
By using the relations in \eqref{params}, the texture-zero relation essentially 
fixes the lightest neutrino mass except for the uncertainty in the values of the 
mixing angles and mass differences.

The solutions are summarized in Table\,\ref{table.results}, where we show 
the possible values for the mass of the lightest neutrino, the effective 
neutrinoless double beta decay parameter ($m_{\beta\beta}$) and the sum of 
neutrino masses.
The mixing angles $\theta_{12},\theta_{13}$ and the mass differences $\Delta 
m^2_{12},\Delta m^2_{23}$ are taken within 3-$\sigma$ of the global fit of 
Ref.\,\cite{lisi} while the values $\theta_{23}=45^\circ$ and $\delta=\pm \pi/2$ 
are fixed by symmetry.

We can see that case III is excluded due to the Planck power spectrum 
limit (95\% C.L.)\,\cite{planck},
\eq{
\sum_{i}m_i< \unit[230]{meV}.
}
We are left with two cases for the normal ordering (NO) and one case for the inverted 
ordering (IO).
All these cases are also compatible with the latest KamLAND-Zen upper limit for the 
neutrinoless double beta decay parameter at 90\%C.L.\,\cite{kamland-zen},
\eq{
m_{\beta\beta}<(61\text{ -- }165)\,\unit{meV}\,.
}
The variation in the latter, comes from the uncertainty in the various evaluations 
of the nuclear matrix elements.
In the future, KamLAND-Zen and EXO-200 experiments will probe the IO region that 
includes our case IV.
\begin{table}[h]
$\begin{array}{|c|c|c|c|c|c|c|}
\hline
\text{Case} & \mfn{(M_\nu)_{\alpha\beta}\!=\!0} & \text{ordering} & \text{CP 
parities} & m_0 
 & m_{\beta\beta} & \sum m_\nu \\
\hline
\text{I} & (ee) & {\rm NO} & (-++) & 4.4\text{ -- }9.0 & 0 & 
    63\text{ -- }74 \\
\text{II} & (ee) & {\rm NO} & (+-+) & 1.1\text{ -- }3.9 & 0 & 
    59\text{ -- }65 \\
\text{III} & (\mu\tau) & {\rm NO} & (++-) & 151\text{ -- }185 & 
    142\text{ -- }178 & 460\text{ -- }561 \\
\text{IV} & (\mu\tau) & {\rm IO} & (+-+) & 15\text{ -- }30 & 
    14.3\text{ -- }29.3 & 116\text{ -- }148 
\\ \hline
\end{array}
$
\caption{\label{table.results}
Possibilities for one-zero textures with predictions for the lightest 
neutrino mass ($m_0$), neutrinoless double beta decay effective mass 
($m_{\beta\beta}$) and sum of neutrino masses; all masses are in \unit{meV}.
}
\end{table}

\begin{figure}[h]
\includegraphics[scale=0.83]{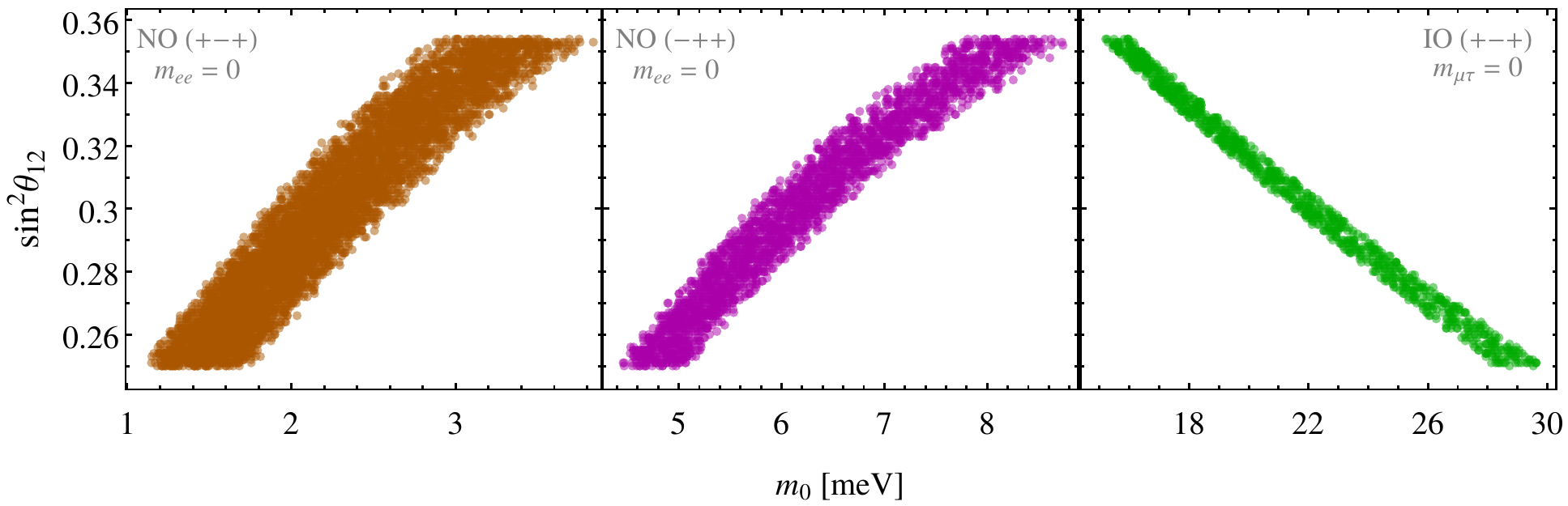}
\caption{
Correlation between $\sin^2\theta_{12}$ and the lightest neutrino mass $m_0$ 
for $\cpmutau$ symmetric neutrino mass matrix with one-zero textures.
The oscillation observables are varied within 3-$\sigma$ of Ref.\,\cite{lisi}
and $m_{\alpha\beta}$ denote $(M_\nu)_{\alpha\beta}$.
}
\label{fig:corr}
\end{figure}

The texture-zero relation $a=0$ or $b=0$ in \eqref{params} also leads to a 
correlation between mixing angles and the lightest neutrino mass when the 
parameters are allowed to vary within the experimental uncertainties.
For the phenomenologically allowed cases, we show this correlation in 
Fig.\,\ref{fig:corr} for $\theta_{12}$.
We can see that the correlation is strong for $\theta_{12}$ while for $\theta_{13}$ 
we have checked that it is only mild.
It is clear that a more precise determination of $\theta_{12}$ will lead to a more
precise prediction for the lightest neutrino mass.
Concerning the neutrinoless double beta decay rates, this information leads to a testable prediction for $m_{\beta\beta}$
in case IV but only to a falsifiable prediction for other cases ($m_{\beta\beta}=0$).

%%%%%%%%%%%%%%%%%%%%%%%%%%%%%%%%%%%%%%%%%%%%%%%%%
\section{Model}
\label{sec:model}

In order to obtain the $\cpmutau$ symmetric mass matrix \eqref{cp.mutau} with vanishing
$(ee)$ or $(\mu\tau)$ entries, it is sufficient to introduce scalars carrying $\ZZ_8$ charges 
corresponding to the nonzero entries in 
\eqref{Z8:mnu}. The absence of appropriate fields will lead to texture-zeros\,\cite{tanimoto}.

Let us introduce SM singlet scalars $\eta_k\sim\omega_8^k$ labelled by their $\ZZ_8$ charges, 
$k\le 7$. We need to know how $\eta_k$ transforms under $\cpmutau$. For that, 
we just need to infer how $\eta_1$ transforms. From \eqref{cp:nu}, we can see that 
the lepton doublet $L_\mu\sim \omega_8$ has the same charge and transforms under $\cpmutau$ as 
\eq{ 
\cpmutau:~~ L_\mu\to L_\tau^{cp}\,, 
} 
where $L_\tau^{cp}\sim \omega_8^{-3}$. So we 
expect 
\eq{ 
\cpmutau:~~ \eta_1\to \eta_{3}^*\,. 
} 
If we double the charge, we 
obtain 
\eq{ 
\label{cp:eta2}
\cpmutau:~~ \eta_2\to \eta_6^*\sim \eta_2\,,
} 
and the last identification can be made if $\eta_2$ carries no other quantum number
besides $\ZZ_8$\,\cite{cp.mutau}. 
Therefore, the fields $\eta_1,\eta_3,\eta_2^*,\eta_2$ couple to the appropriate 
quadratic combination in \eqref{Z8:mnu} giving rise to the $(e\tau), (e\mu), (\mu\mu)$ and $(\tau\tau)$
entries of $M_\nu$, respectively.

To avoid the $(ee)$ combination in \eqref{Z8:mnu} to acquire a bare coupling,% 
\footnote{In the case of $(M_\nu)_{\mu\tau}=0$ we can obtain $(M_\nu)_{ee}$ from 
the bare coupling but it will come from a operator which is lower order than the rest.} 
we introduce a $\ZZ_4^{B-L}$ symmetry under which the leptons have 
charge $-i$ while the scalar $\eta_k\sim -1$.%
\footnote{Since this charge is real, the identification of $\eta_6^*=\eta_2$ in
\eqref{cp:eta2} is consistent.\,\cite{cp.mutau}}
If we also introduce the \textit{real} fields $\eta_0$ and $\eta_4$, the dimension five
Weinberg operator will come from the nonrenormalizable operators
\eqali{
&\ums{2}\frac{c_{ee}}{\Lambda^2}\eta_0 L_eHL_eH
+\frac{c_{\mu\tau}}{\Lambda^2}\eta_4 L_\mu HL_\tau H
%\cr
+\frac{c_{e\mu}}{\Lambda^2}\eta_3 L_eHL_\mu H 
+\frac{c_{e\tau}}{\Lambda^2}\eta_1 L_eHL_\tau H 
\cr&
+\ \ums{2}\frac{c_{\mu\mu}}{\Lambda^2}\eta_2^* L_\mu HL_\mu H 
+\ums{2}\frac{c_{\tau\tau}}{\Lambda^2}\eta_2 L_\tau HL_\tau H 
+h.c.
}
Since $\cpmutau$ symmetry ensures $c_{e\tau}=c_{e\mu}^*$, 
$c_{\tau\tau}=c_{\mu\mu}^*$ and real $c_{ee},c_{\mu\tau}$, we obtain the $\cpmutau$ 
symmetric mass matrix \eqref{cp.mutau} if $\cpmutau$ is not broken by $\eta_k$, i.e.,
\eq{
\label{vevs}
\aver{\eta_3}=\aver{\eta_1}^*\,.
}
We show in the following that these symmetric vevs are possible.

Finally, the texture-zero in the $(ee)$ or $(\mu\tau)$ entry follows if $\eta_0$ or $\eta_4$ 
is absent.
In this effective case, the texture-zero is not exact because even if $\eta_4$ is absent it can be replaced by 
e.g. $\eta_0\eta_2^2$ or $\eta_1^*\eta_3\eta_2$ but it only appears with three $\eta_k$ fields due to 
$\ZZ_4^{B-L}$ and give entries in the neutrino mass matrix 
suppressed by $\aver{\eta_k}^2/\Lambda^2$.
Possibly, this suppression can be improved in a specific UV complete model.
Some examples of UV completions for the case where the abelian symmetry is 
$\mathtt{L}_\mu-\mathtt{L}_\tau$ can be seen in Ref.\,\cite{cp.mutau}.

The remaining task is to check that the scalar potential involving 
$\eta_k$ can be minimized by values conserving $\cpmutau$, i.e., obeying \eqref{vevs}.
The potential contains no trilinear terms and the terms that depend on their 
phases are only quartic:
\eqali{
V\supset &~
\lambda_1(\eta_1\eta_3)^2
+\lambda_1'\eta_1^*\eta_3(\eta_1^{*2}+\eta_3^2)
+\lambda_2\eta_2^4
+\lambda_3\eta_2^2(\eta_1\eta_3+\eta_1^*\eta_3^*)
\cr& 
+\lambda_4\eta_0\eta_2^*\eta_1^*\eta_3
+\lambda_4'\eta_0\eta_2(\eta_1^{*2}+\eta_3^2)
+h.c.,
}
where $\lambda_1$ is real while the rest are complex.
This corresponds to the case where $\eta_4$ is absent.

If we parametrize $\aver{\eta_k}=u_ke^{i\alpha_k}$, we can see that the only 
terms that depend on the combination $\alpha_{13}\equiv \alpha_1+\alpha_3$ are the 
terms with coefficients $\lambda_1$, $\lambda_1',\lambda_3,\lambda_4'$.
The $\cpmutau$ symmetry corresponds to
\eq{
u_1\leftrightarrow u_3\,,\quad
\alpha_1\leftrightarrow -\alpha_3\,,
}
which flips the sign of $\alpha_{13}$ while the rest are invariant.
For the terms with $\lambda_1$ and $\lambda_3$, the dependence is through 
$\cos2\alpha_{13}$ and $\cos\alpha_{13}$ respectively.
The dependence for the terms with $\lambda_1'$ and $\lambda_4'$ is only through 
$(\eta_1^{*2}+\eta_3^2)$ that have the form
\eq{
u_1^2\cos(\alpha_{13}+\varphi)+u_3^2\cos(\alpha_{13}-\varphi)\,.
}
This expression depends on $\alpha_{13}$ only through $\cos\alpha_{13}$ if 
$u_1=u_3$.
In this case, for all these terms, parameters can be chosen so that $\alpha_{13}=0$ 
is a minimum.

To check that $u_1=u_3$ can be achieved, we can gather the terms that do not depend 
on $\alpha_{13}$ and write the quadratic contributions for $u_1$ and $u_3$, after 
taking the minimizing values for other parameters, as
\eq{
\label{quadratic}
A(u_1^2+u_3^2)+2Bu_1u_3\,,
}
where $B$ comes from the $\lambda_3,\lambda_4$ terms.
If we arrange $A+B<0$ and $A-B>0$, $u_1-u_3=0$ minimizes the quadratic terms in 
\eqref{quadratic} and hence the whole potential in the direction orthogonal to 
$(u_1,u_3)\sim (1,1)$.
This checks that the symmetric minimum \eqref{vevs} is possible.
We have also checked that numerically it is easy to obtain the symmetric minimum.

If $\eta_4$ is present instead of $\eta_1$, the terms with coefficients $\lambda_4$ 
and $\lambda_4'$ are replaced by
\eq{
\lambda_4\eta_2\eta_4\eta_1^*\eta_3
+\lambda_4'\eta_2^*\eta_4(\eta_1^{*2}+\eta_3^{2})
+h.c.
}
We arrive at the same result as before: there are parameter regions where 
$\cpmutau$ remains conserved by the vevs.

%%%%%%%%%%%%%%%%%%%%%%%%%%%
\section{Conclusions}
\label{sec:conclusion}

We have shown by explicit construction a highly predictive scenario where the 
neutrino mass matrix is symmetric by $\cpmutau$ or $\mu\tau$-reflection and 
\textit{additionally} contains one texture-zero in the $(ee)$ or $(\mu\tau)$ entry. 
Besides the usual predictions of $\cpmutau$ -- maximal $\theta_{23}$, maximal Dirac 
CP phase and trivial Majorana phases -- we find that only two values for $m_1$ are 
possible for normal ordering and only one value for $m_3$ is possible for the 
inverted ordering. The NO solutions correspond to $m_1$ of a few 
meV and the IO solution has $m_3$ of around 20 meV. The specific intervals when we 
allow for the uncertainties in the oscillation parameters can be seen in 
Table\,\ref{table.results} together with the possible CP parities, the value for 
the neutrinoless double beta decay parameter and the sum of neutrino masses. The 
strong correlation that appears between the solar angle $\theta_{12}$ and the 
lightest neutrino mass is shown in Fig.\,\ref{fig:corr}. 
The IO solution is expected to be tested in the near future by the neutrinoless double beta decay 
experiments such as KamLAND-Zen and EXO-200 as they reach the IO region.
For the solutions with NO, we predict a negligible neutrinoless double beta decay 
rate as
$m_{\beta\beta}\approx 0$ which can be falsified but will be impossible to confirm.
Finally, the possibility of a neutrino mass matrix with $\cpmutau$ symmetry 
\textit{simultaneously} with a texture-zero that is enforced by symmetry 
was first shown here and it is only allowed by combining in a 
non-usual way a discrete abelian symmetry at least as large as $\ZZ_8$ and 
$\cpmutau$.

%%%%%%%%%%%%%%%%%%%%%%%%%%%
\acknowledgements

C.C.N acknowledges partial support by Brazilian Fapesp grant 2013/26371-5 and 
2013/22079-8. 
B.L.S.V. would like to thank Coordenação de Aperfeiçoamento de Pessoal de Nível 
Superior (CAPES), Brazil, for financial support.

%%%%%%%%%%%%%%%%%%%%%%%%%%%

%%%%%%%%%%%%%%%%%%%%%%%%%%%%%%%%%%%%%%%%%%%%%%%%%
\end{document}